\documentclass[aps,prb,floatfix,twocolumn,superscriptaddress,showpacs,amsmath,amssymb]{revtex4}
\usepackage{graphicx}
\usepackage{epsfig} 
\usepackage{amssymb}
\usepackage{amsmath}
\usepackage{color}
\allowdisplaybreaks

\begin{document}
\title{Molecular bonding with the RPAx: from weak dispersion forces to\\ strong correlation} 

\author{Nicola Colonna}
\affiliation{International School for Advanced Studies (SISSA), Via Bonomea 265, I-34136 Trieste, Italy}
\affiliation{Theory and Simulation of Materials (THEOS), \'Ecole Polytechnique F\' ed\'erale de Lausanne, 1015 Lausanne, Switzerland} 
\author{Maria Hellgren}
\affiliation{International School for Advanced Studies (SISSA), Via Bonomea 265, I-34136 Trieste, Italy}
\affiliation{Physics and Materials Science Research Unit, University of Luxembourg, 162a avenue de la Fa\"{\i}encerie, L-1511 Luxembourg, Luxembourg} 
\affiliation{IMPMC, Sorbonne Universit\'es,                                                                                 
UMR CNRS 7590, Mus\'eum National d'Histoire Naturelle, IRD UMR 206, 4 Place Jussieu, F-75005 Paris, France} 
\author{Stefano de Gironcoli}
\affiliation{International School for Advanced Studies (SISSA), Via Bonomea 265, I-34136 Trieste, Italy}
 \affiliation{CRS Democritos, CNR-IOM Democritos, Via Bonomea 265, I-34136 Trieste, Italy}

\date{\today} 


\begin{abstract}
In a recent article [Phys. Rev. B 90, 125102 (2014)] we showed that the random phase approximation with 
exchange (RPAx) gives accurate total energies for a diverse set of systems including 
the high and low density regime of the homogeneous electron gas, the N$_2$ molecule, and the 
H$_2$ molecule at dissociation. 
In this work, we present results for the van der Waals bonded Ar$_2$ and Kr$_2$ dimers and demonstrate 
that the RPAx gives superior dispersion forces as compared to the RPA. We then show that this improved 
description is crucial for the bond formation of the Mg$_2$ molecule. In addition, 
the RPAx performs better for the Be$_2$ dissociation curve at large nuclear separation but, similar to the RPA, 
fails around equilibrium due to the build up of a large repulsion hump. For the strongly correlated LiH 
molecule at dissociation we have also calculated the RPAx potential and find that the correlation peak at 
the bond mid-point is overestimated as compared to the RPA and the exact result. The step feature is 
missing and hence the delocalisation error is comparable to the RPA. This is further illustrated by a 
smooth energy versus fractional charge curve and a poor description of the LiH dipole moment at dissociation.   
\end{abstract}
\maketitle

\section{Introduction}\label{sect:sect0}

In its formulation due to Kohn and Sham (KS),~\cite{kohn_self-consistent_1965} Density Functional Theory 
(DFT)~\cite{hohenberg_inhomogeneous_1964} allows the computation of the ground state density and total energy 
of a many-particle system by solving a set of single particle Schr\"odinger-like equations.
Within current approximations to the unknown exchange-correlation (xc) energy, this can be done
with a computational cost that grows as a power of the number of particles, thus making 
KS-DFT very attractive from a computational point of view and the ideal tool to study complex systems.
The local spin density approximation (LSDA)~\cite{kohn_self-consistent_1965,gunnarsson_exchange_1976} and its generalized 
gradient corrections (GGAs)~\cite{becke_density-functional_1988,perdew_generalized_1996} are the 
most widely used xc energy functionals. Notwithstanding their capability to successfully predict the properties of a wide class 
of electronic systems, there are still many situations in which they perform poorly or even
fail qualitatively. One such fundamental problem is the description of dispersion interactions.
Although one of the weakest, the dispersion interaction is ubiquitous and can add up to 
give a substantial contribution to the interaction energy in large assemblies of atoms and molecules
thus becoming very important in systems like, for instance, biomolecules, adsorbates or nanomaterials.
Different solutions have been proposed to deal with weakly interacting systems in DFT, ranging 
from semi-empirical approaches~\cite{grimme_consistent_2010,tkatchenko_accurate_2009,silvestrelli_van_2008} to truly
non-local and parameter-free functionals.~\cite{dion_van_2004,thonhauser_van_2007,vydrov_nonlocal_2010,lee_higher-accuracy_2010}

Another class of materials for which standard Density Functional Approximations (DFAs) badly fail 
is represented by the strongly correlated systems.
Strong correlation is usually meant to refer to situations where multiple determinants associated
with degeneracy or near degeneracy are needed and a single-particle picture becomes inadequate.
This limitation leads, for instance, to a dramatic failure of standard KS methods when studying 
the dissociation of systems into open shell fragments, whose simplest and paradigmatic example is
the stretched hydrogen molecule.
However, this is not a failure of DFT itself, but just a break-down of present DFAs. 
In this respect many different corrective approaches have been developed for \emph{ab initio} 
calculations on strongly correlated systems ranging 
from Self Interaction Correction~\cite{perdew_self-interaction_1981} to Exact-Exchange and hybrid 
functionals~\cite{heyd_hybrid_2003} to DFT+U.~\cite{anisimov_band_1991}

However, rather than introducing \emph{ad hoc} corrections to describe selected problems,
it would be highly desirable to find a new class of functionals that perform
uniformly better than the present ones as already happened passing from LDA to GGAs and more recently 
with the development of hybrid functionals. In this respect it is illuminating to start with the formally exact 
way of constructing the exchange and correlation functional using the Adiabatic Connection Fluctuation-Dissipation
(ACFD) theory.~\cite{langreth_exchange-correlation_1975,gunnarsson_exchange_1976,langreth_exchange-correlation_1977}
In this framework an exact expression for the exchange-correlation energy can be derived in terms
of the density-density response functions of a continuous set of fictitious systems defining a path that couples 
the non-interacting KS system with the real many-body interacting one.
The formalism thus provides a powerful theoretical framework for a systematic development of advanced 
functionals~\cite{colonna_correlation_2014} but also a practical way to compute accurate correlation energies. 
Moreover, all ACFD 
methods treat the exchange energy exactly thus cancelling out the spurious self-interaction error present 
in the Hartree energy, while the correlation energy is fully non-local and automatically includes van der Waals 
interactions. 

Although much more expensive than the simple approximations to it, a direct evaluation of the exact formula
is also possible when the interacting density-density response function is given. A Dyson-like equation
linking the latter to its non-interacting counterpart can be derived from time dependent density functional 
theory (TDDFT), but still requires a suitable approximation for the unknown exchange-correlation kernel, 
$f_{\rm xc}$.\cite{runge_density-functional_1984,gross_local_1985} Nevertheless it has been shown that using the 
Random Phase Approximation (RPA), i.e., by completely neglecting the xc kernel, plus a local-density correction for 
short range correlation, leads to a qualitatively correct description
of van der Waals interactions,~\cite{fuchs_accurate_2002, marini_first-principles_2006, harl_cohesive_2008, nguyen_first-principles_2010} static correlation~\cite{furche_molecular_2001,fuchs_describing_2005} and systems with mixed bonds.\cite{olsen_rpa2011,olsen_rpa2013}

On the other hand, the RPA total energy is far too negative~\cite{jiang_random-phase-approximation-based_2007,hellgren_correlation_2007} 
because of the well know overestimation of the correlation energy;~ \cite{kurth_density-functional_1999} moreover, the 
binding energy of van der Waals systems is systematically underestimated and an erroneous repulsion
hump appears in the dissociation curve of covalently bonded systems at intermediate distances.
For an exhaustive review on the RPA see for instance Refs.~\onlinecite{ren_random-phase_2012,Eshuis2012} 
and citations therein.

Staying within the ACFD framework, a rigorous possibility to address the shortcomings of RPA is to 
include a xc contribution to the kernel that appears in the Dyson-like equation defining the 
interacting density-density response function. For a consistent description to first order in the 
electron-electron interaction, not only the Coulomb kernel, defining RPA, 
but also an exchange contribution has to be taken into account. 
The frequency-dependent exact-exchange kernel, $f_{\rm x}$, has been derived
by G\"orling~\cite{gorling_exact_1998,gorling_exact_1998_2,kim_exact_2002} from the time-dependent 
optimized effective potential (TDOEP) method, by Hellgren and von 
Barth~\cite{hellgren_linear_2008,hellgren_exact-exchange_2009} from a variational formulation 
of many-body perturbation theory (MBPT) and more recently by us~\cite{colonna_correlation_2014}
from a perturbative approach along the adiabatic-connection path. 
The corresponding approximation for the density-density response function, 
named RPAx, is obtained by solving the Dyson equation approximating the 
xc kernel, $f_{\rm xc}$, with its exact-exchange contribution, $f_{\rm x}$, 
and has been successfully used in the ACFD formula to compute 
correlation energies of atoms~\cite{hellgren_linear_2008, hellgren_correlation_2010} and
molecules.~\cite{hesselmann_random_2010,heselmann_correct_2011,bleiziffer_resolution_2012,
bleiziffer_self-consistent_2015}
In a recent work by us we showed that with a slight redefinition 
of the re-summation procedure one can overcome a severe deficiency of the 
RPAx when it is applied to the low density regime of the electron gas and 
the stretched N$_2$ dimer.\cite{colonna_correlation_2014}
 
In this work, we investigate the RPAx for molecular dissociation of diatomic 
molecules with different types of chemical bonds. The binding energies as 
well as the entire potential energy surfaces are obtained. We study the weakly 
interacting Kr$_2$ and Ar$_2$ dimers which allows us to test how well the 
van der Waals forces are described. We also study the Be$_2$ and Mg$_2$ dimers 
which are challenging due to the complex interplay between van der Waals 
and covalent bonding forces. Finally, we address the question of how well the 
RPAx can describe strong correlation by analyzing the LiH molecule. 

The paper is organized as follows. Section~\ref{sect:theory} 
briefly summarizes the basic equations of the ACFD theory, the RPAx approximation 
and the computational details of our calculations. Results for dissociation curves 
and structural parameters of different kind of molecules are reported and discussed 
in Sec.~\ref{sect:res}. In Sec.~\ref{sect:summ} we give our conclusions.

\section{Theory}\label{sect:theory}

In this section we recall some basic ideas concerning the ACFD theory 
as well as some technical aspects of the plane-wave and pseudopotential implementation
of the RPA(x) total energy.

\subsection{Exchange and correlation energy within ACFD theory}

Within the ACFD framework a formally exact expression for the 
exchange-correlation energy $E_{\rm xc}$ of an electronic system can 
be derived:~\cite{langreth_exchange-correlation_1975,langreth_exchange-correlation_1977}
\begin{align}
  E_{\rm xc}= & -\frac{1}{2} \int_0^1 d\lambda \int d\mathbf{r} d\mathbf{r}' \frac{e^2}{|\mathbf{r}-\mathbf{r}'|} 
 \nonumber \\
 & \times \left\{ \frac{\hbar}{\pi}\int_0^{\infty}\!\!\!\! \chi_{\lambda}(\mathbf{r},\mathbf{r}' ; iu) \; du \;\; +\; \delta(\mathbf{r}-\mathbf{r}')
   n(\mathbf{r})  \right\}
 \label{eq1.1}
\end{align}
where $\chi_{\lambda}(\mathbf{r},\mathbf{r}' ; iu)$ is the density-density response 
function at imaginary frequency, $iu$, of a system whose electrons interact 
with a scaled Coulomb interaction, $\lambda e^2/|\mathbf{r}-\mathbf{r}'|$, 
and move in a local potential chosen in such a way to keep the electronic 
density fixed to the ground state density of the fully interacting system 
($\lambda =1$).
At $\lambda = 1$, the local potential is equal to the external potential 
(usually the nuclear potential) of the fully interacting system and 
$H_{\lambda=1}$ coincides with the fully interacting Hamiltonian while 
at $\lambda = 0$ the local potential coincides with the KS potential 
and $H_{\lambda=0}$ is the KS Hamiltonian.

The exchange-correlation energy can be further separated into the KS 
exact-exchange energy, E$^{\rm EXX}_{\rm x}$ and the correlation energy E$_{\rm c}$. 
The former has the same expression as the Hartree-Fock exchange energy but is 
evaluated with the KS orbitals $\phi_i(\mathbf{r})$,
\begin{equation}
 E_{\rm x}^{\rm EXX} = -\frac{e^2}{2}\int d\mathbf{r} d\mathbf{r}' \frac{|\sum_i^{occ} \phi_i^*(\mathbf{r}) \phi_i(\mathbf{r}')|^2}
 {|\mathbf{r}-\mathbf{r}'|}.
 \label{eq1.2bis}
\end{equation}
It is easy to verify that it can be derived from Eq.~(\ref{eq1.1}) 
replacing $\chi_{\lambda}$ with the non-interacting density response 
function
\begin{equation}
\chi_0(\mathbf{r},\mathbf{r}';iu) = \sum_{ij}(f_i-f_j)
\frac{\phi_i^*(\mathbf{r}) \phi_j(\mathbf{r}) \phi_j^*(\mathbf{r}') \phi_i(\mathbf{r}')}
     {\epsilon_i-\epsilon_j + i\hbar u} 
 \label{eq1.2ter}
\end{equation}
where $\epsilon_i$, $\phi_i(\mathbf{r})$ and $f_i$ are the KS eigenvalues, KS orbitals
and occupation numbers, respectively.  
Subtracting the KS exchange energy from Eq.~(\ref{eq1.1}) 
the correlation energy $E_{\rm c}$ is obtained in terms of linear density-density
responses:
\begin{equation}
 E_{\rm c} = -\frac{\hbar}{2\pi} \int_0^1 d\lambda \int_0^{\infty} du\; \text{Tr} \left\{ \upsilon_{\rm c} \left[ \chi_{\lambda}(iu) 
 - \chi_0(iu) \right] \right\} 
  \label{eq1.3}
\end{equation}
where $\upsilon_{\rm c}=e^2/|\mathbf{r}-\mathbf{r}'|$ is the Coulomb kernel and Tr indicates the integrals over spatial coordinates. 

For $\lambda > 0$ the interacting density response function 
$\chi_{\lambda}(iu)$ can exactly be related to the KS non-interacting one 
via a Dyson equation obtained from time-dependent density functional theory (TDDFT):
\begin{equation}
 \chi_{\lambda}(iu) = \chi_0(iu) + \chi_0(iu) [ \lambda \upsilon_{\rm c} + f^{\lambda}_{\rm xc}(iu)] \chi_{\lambda}(iu)
\label{eq1.4}
\end{equation}
where $f^{\lambda}_{\rm xc}(iu)$ is the scaled frequency-dependent xc kernel. 
Spatial coordinate dependence is implicit in the matrix notation.
When the xc kernel is specified one can thus determine a corresponding 
correlation energy via Eq. (\ref{eq1.3}). 
Alternatively, the interacting density response function can be obtained from 
many-body perturbation theory which allows truncation at arbitrary order
and various re-summations of the higher order terms. For example, by truncating to 
first order we generate an approximation similar to the MP2 (Moller-Plesset second order perturbation theory) 
approximation.\cite{mp2jiangengel,hellgren_correlation_2007} A different RPAx re-summation based on a first order expansion of the irreducible polarisability was used in our previous work to eliminate a 
pathology in the original formulation of RPAx.\cite{colonna_correlation_2014} 
 
\subsection{RPAx energy functional}

The RPA and RPAx correlation energies $E_{\rm c}^{\rm RPA(x)}$ are obtained by
solving the Dyson equation~(\ref{eq1.4}) setting $f^{\lambda}_{\rm xc} = 0$ and
$f^{\lambda}_{\rm xc} = \lambda f_{\rm x}$, respectively, and inserting the result in 
the ACFD formula. The RPA(x) total energy is then defined as
\begin{align}
 E^{\rm RPA(x)} = & T_{\rm s} + E_{\rm ext} + E_{\rm H} + E_{\rm x}^{\rm EXX} + E_{\rm c}^{\rm RPA(x)}
\label{eq:total-ene}
\end{align}
where $T_{\rm s}$, $E_{\rm ext}$ and $E_{\rm H}$ are the KS kinetic, external and Hartree energy, 
respectively. 

In this work we have also exploited alternative re-summations\cite{colonna_correlation_2014,RPAx-F} based on the RPAx defined as
\begin{equation}
 \chi^{\rm tRPAx}_{\lambda} =  P_{\lambda}^{(1)} + \lambda P_{\lambda}^{(1)} \upsilon_{\rm c} \chi^{\rm tRPAx}_{\lambda}
 \label{eq3.4}
\end{equation}
with $ P_{\lambda}^{(1)} = \chi_0 + \lambda \chi_0 f_{\rm x} \chi_0$ and
\begin{equation}
 \chi^{\rm t^{\prime}RPAx}_{\lambda} = \chi_{\lambda}^{\rm RPA} + \lambda \chi_{\lambda}^{\rm RPA} f_{\rm x} \chi_{\lambda}^{\rm RPA}.
\label{eq3.5}
\end{equation}
Both tRPAx and t'RPAx have shown to give results either very similar to the original RPAx or results which 
qualitatively improve upon the RPAx.
In particular, we observed that the break-down of the RPAx approximation 
in the low density regime of the homogeneous electron gas and in the stretched nitrogen dimer is lifted by using the 
tRPAx and t'RPAx resummations~\cite{colonna_correlation_2014}. The difference between the RPAx and the alternative resummations for $\chi$
lies in the treatment of the terms beyond first order in the Coulomb interaction. Crucial to the original 
RPAx is the so-called exchange kernel which is defined as the functional derivative of the exchange OEP potential.
When inserted into the linear response equation it can be shown to exactly reproduce first order 
particle-hole and self-energy effects as defined within linear response 
time-dependent Hartree-Fock (TDHF) theory.\cite{hellgren_linear_2008} However, higher orders cannot be 
represented in terms of higher order particle-hole and self-energy diagrams and therefore a direct 
comparison between the RPAx- and the TDHF response function cannot be made. On the other hand, it is 
well known that TDHF produces a particle-hole interaction in solids that is too strong.\cite{bruneval_beyond_2006} 
This is to some extent compensated by the HF band structure which overestimates the quasi-particle gap. 
It can therefore be argued that the instabilities of the RPAx response function that are not present in the 
TDHF response function are due to the fact that to higher orders the exchange kernel is not able to correctly 
balance the particle-hole attraction with a DFT band structure that has an energy gap much smaller 
than the HF band structure. The use of tRPAx and t'RPAx removes this type of instability by simply neglecting 
higher order particle-hole terms in the response function. In the cases where RPAx is well-behaved we 
found that this procedure has only a minor effect on the total energies.

In this work we focus on systems with a rather large gap (or on two-electron systems for which the exact-exchange kernel 
simply reduces to a half of the Hartree kernel) and we do not expect any critical behavior of the original RPAx. 
Nevertheless we tested all the alternative resummations for all the systems considered in this work finding 
additional cases (Be$_2$ and Mg$_2$ which have a relatively small gap around equilibrium) where the 
alternative resummations perform quantitatively better than the original RPAx.

 \begin{figure}[t]
 \begin{center}
 \includegraphics[scale=0.3]{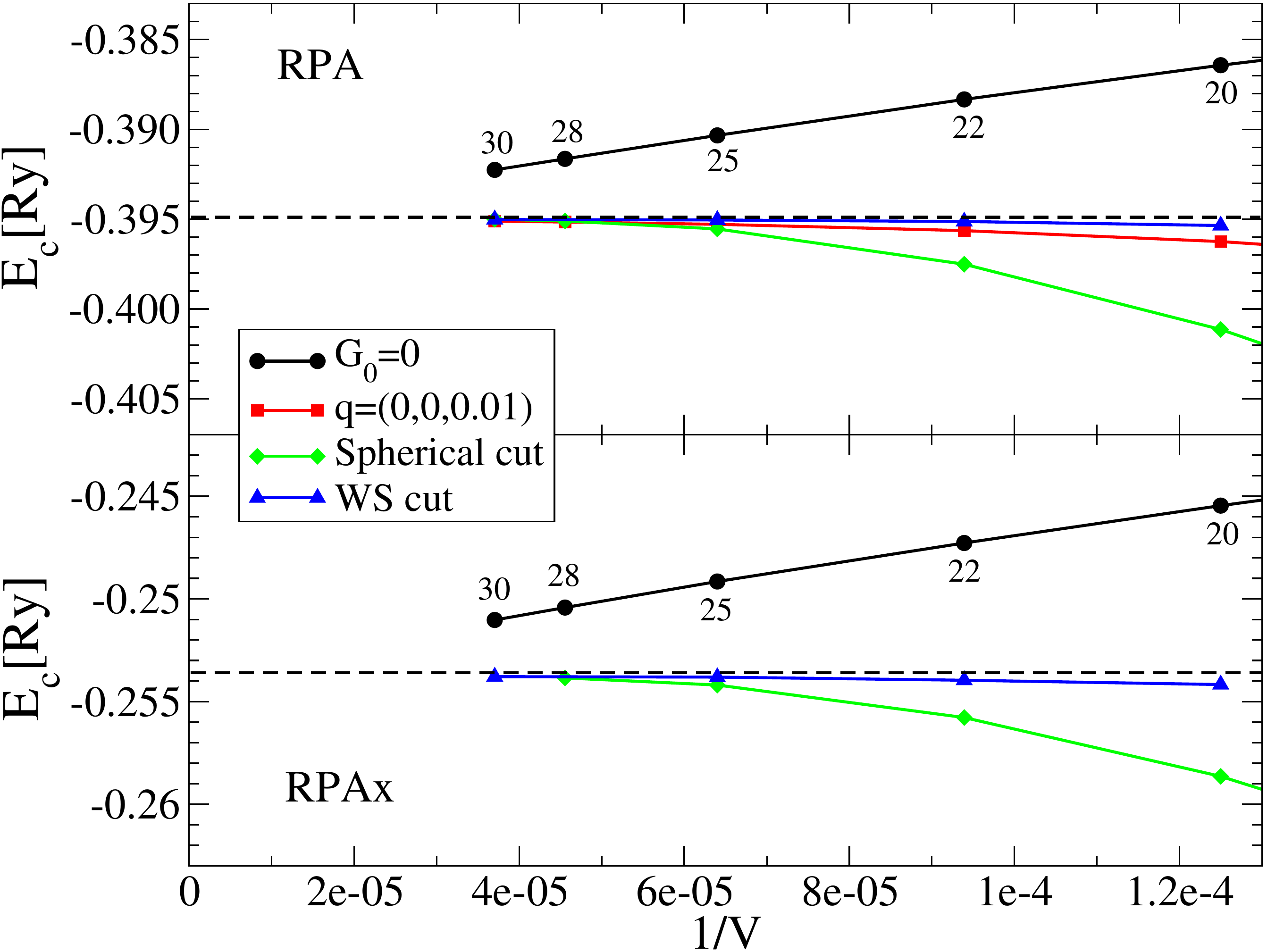}
 \caption{Convergence of the RPA (top panel) and RPAx (bottom panel) correlation energy $E_c$ of the Be$_2$ molecule at
an inter-atomic distance $d=2.45$ \AA\ , with respect to the size of the super-cell. The numbers close to the black
circles indicate the linear size in bohr of the cubic super-cell. Different schemes for treating the
$\mathbf{q}\rightarrow\mathbf{0}$ limit are compared. See text for details.}
 \label{fig:conv-Ec-alat}
 \end{center}
 \end{figure}
 In principle a fully self consistent field (scf) evaluation of the RPA(x) energy can be done
by relying on the optimized effective potential (OEP) method which provides a way to compute the local 
multiplicative potential that minimizes a given orbital dependent energy functional. 
The first approximate RPA-potential calculations were performed on bulk materials.\cite{godby_accurate_1986,godby_self-energy_1988,
kotani_optimized-effective-potential_1998, gruning_effect_2006,gruning_density_2006}
Fully scf-RPA calculations have been performed for closed shell atoms~\cite{hellgren_correlation_2007, 
verma_increasing_2012} and simple molecules.~\cite{hellgren_correlation_2012, caruso_bond_2013, bleiziffer_efficient_2013}
For what concerns plane wave implementations, an efficient way to compute the scf-RPA energy and potential has
been recently proposed by Nguyen \emph{et al.}~\cite{nguyen_ab_2014}, stemming from the non-scf implementation
of Nguyen and de Gironcoli~\cite{nguyen_efficient_2009} paving the way for extensive calculations also for 
extended systems. The functional derivative of the RPAx correlation energy was first calculated for spherical
atoms~\cite{hellgren_correlation_2010} and recently for molecules.~\cite{bleiziffer_self-consistent_2015}
However, despite the efforts in trying to reduce the computational workload of RPA and RPAx calculations, 
they still remain computationally very demanding. For this reason, most of the RPA/RPAx calculations are 
limited to a post self-consistent correction where the xc energy is computed from the charge density obtained 
from a self-consistent calculation performed with a more traditional xc functional. 

The calculations of the RPA and RPAx correlation energies presented in this work are based on an eigenvalue
decomposition of the non-interacting response function and of its first order correction in the limit of vanishing
electron-electron interaction, respectively. Only a small number of eigenvalues contributes significantly to the
decomposition meaning that accurate results for the RPA and RPAx correlation energy can be obtained considering only
the most relevant ones. A detailed and general (base-independent) description of the schemes can be found in
Ref.~\onlinecite{nguyen_efficient_2009} and Ref.~\onlinecite{colonna_correlation_2014}. In the plane-wave
pseudopotential approach any system is treated as periodic and the response functions are block diagonal and can be classified by a vector
$\mathbf{q}$ in the first Brillouin zone. The final expression of the RPAx correlation energy (Eq. (12) in
Ref.~\onlinecite{colonna_correlation_2014}) in a plane-wave basis set becomes: 
\begin{align}
   E_{c}^{\rm RPAx} = -\frac{\hbar}{2\pi} \int_0^{\infty}& du\; \frac{1}{N_q}\sum_{q=1}^{N_q} 
 \sum_{\alpha} \frac{s_{\alpha}(\mathbf{q} ,iu)}{a_{\alpha}(\mathbf{q};iu)} \times \nonumber \\
 & \times \big\{ a_{\alpha}(\mathbf{q};iu) + \ln[1-a_{\alpha}(\mathbf{q};iu)]\big\}
\label{eq:ec-RPAx-PW}
\end{align}
where  $s_{\alpha}(\mathbf{q},iu) = \langle \omega^{\mathbf{q}}_{\alpha} | \chi^{\mathbf{q}}_0(iu) \upsilon_{\rm c}
\chi^{\mathbf{q}}_0(iu) | \omega^{\mathbf{q}}_{\alpha} \rangle $ 
and $\left\{a_{\alpha}(\mathbf{q};iu), | \omega^{\mathbf{q}}_{\alpha} \rangle\right\}$ are the eigenpairs solution of
the generalized eigenvalue problem 
\begin{equation}
 -h_{Hx}^{\mathbf{q}}(iu) | \omega^{\mathbf{q}}_{\alpha} \rangle = a_{\alpha}(\mathbf{q};iu) [-\chi^{\mathbf{q}}_0(iu)]
|\omega^{\mathbf{q}}_{\alpha} \rangle,
\label{eq:RPAx-eig-PW}
\end{equation}
where $h_{Hx}^{\mathbf{q}}(iu) = \chi^{\mathbf{q}}_0(iu)[  v(\mathbf{q} ) + f_x(\mathbf{q};iu)] \chi^{\mathbf{q}}_0(iu)$. \\
When the exact exchange kernel is neglected the RPA energy is recovered~\cite{nguyen_efficient_2009}.

\begin{figure}[t]
\begin{center}
\includegraphics[scale=0.305]{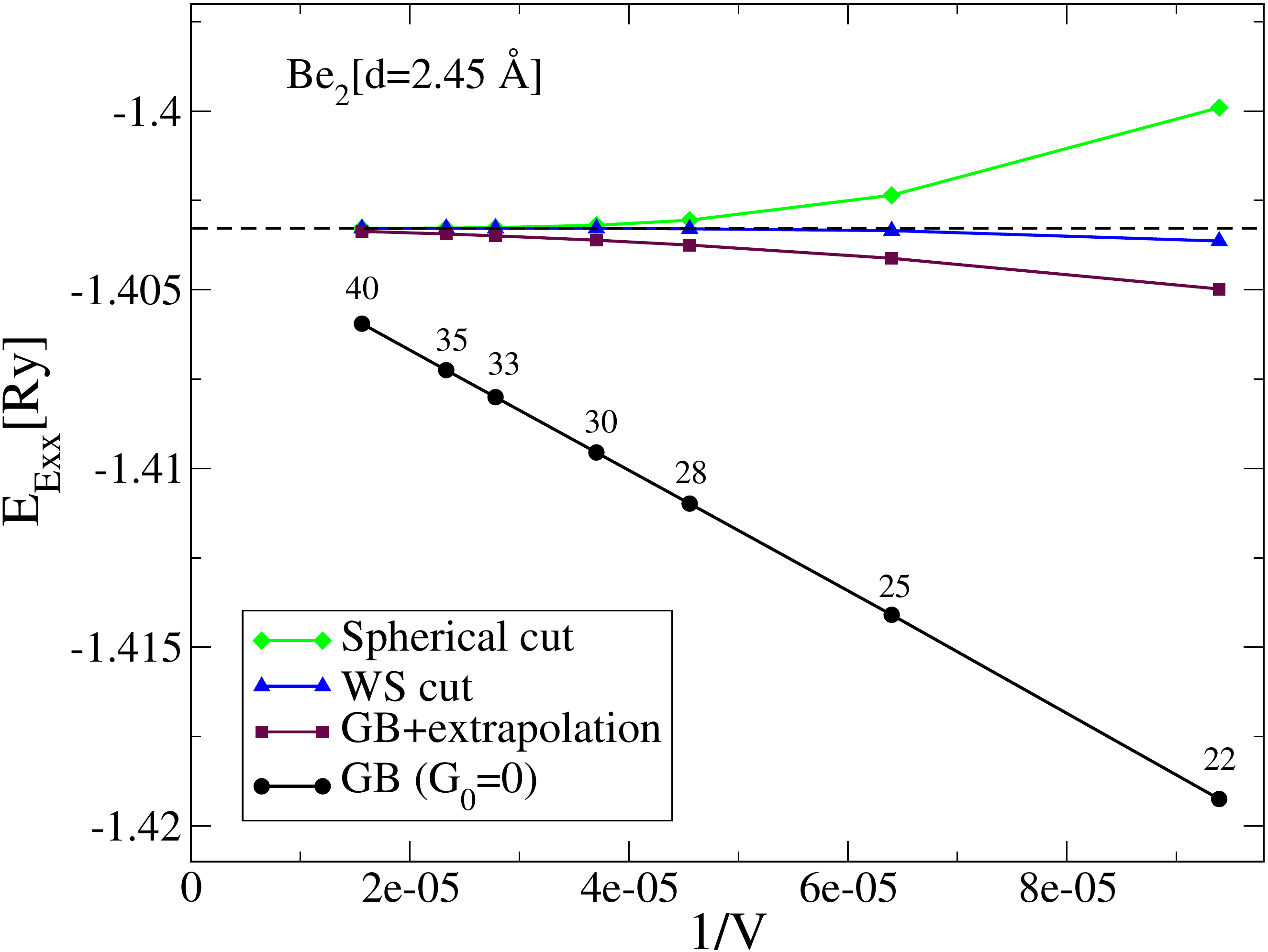}
\caption{Convergence of exact-exchange energy $E_x$ of the Be$_2$ molecule at an inter-atomic distance 
$d=2.45$ \AA\ , with respect to the size of the super-cell. The numbers close to the black circles indicate the linear size in
bohr of the cubic super-cell. Different schemes for treating the integrable divergence in the exact-exchange expression
are compared. See text for details.} 
\label{fig:conv-Ex-alat}
\end{center}
\end{figure}
The dependence of the RPA correlation energy on the Brillouin-zone sampling and super-cell size has been
already carefully analyzed by Harl \emph{et al.}\cite{harl_cohesive_2008} and Nguyen and de Gironcoli~\cite{nguyen_efficient_2009},
with particular emphasis on how to treat the singularity of the Coulomb kernel in the $\mathbf{q}\rightarrow 0$ limit.
In the special case of finite systems the singularity problem can be easily avoided using a modified coulomb interaction. 
The long range tail of the coulomb interaction in real space is truncated after a distance of the order 
of half of the super-cell size making its Fourier transform remaining finite for $\mathbf{q} \rightarrow 0$
thus leading to a well defined problem even exactly at $\mathbf{q}=0$.
The approximation becomes exact in the limit of infinite super-cell meaning 
that a careful check of the convergence with respect to the super-cell size is needed. In Fig.~\ref{fig:conv-Ec-alat}
we report the result for the RPA and RPAx correlation energies of Be dimer at equilibrium distance as a function of
the volume of the simulation box calculated within different approximations.
Simply setting to zero the $|\mathbf{q}+\mathbf{G}|=0$ component ($G_0=0$ in the figure) leads to a slowly convergent 
behavior for the correlation energies that eventually goes to a limiting value (extrapolated with a linear fit of 
the $G_0=0$ data) represented by the dashed black lines in the figure. For the RPA correlation 
energy (top panel) we also report the result from a calculation with a shifted $\mathbf{q}$ mesh (in particular we
set $\mathbf{q}=(0,0,0.01)$) and notice that a much faster convergence toward the same limit of the $G_0=0$
calculation is obtained.
We then tested two different modified Coulomb interactions 
implemented in the {\sc Quantum ESPRESSO} distribution~\cite{giannozzi_quantum_2009}. 
The first one, referred to as ``Spherical cut'' in the figure, is an abrupt truncation of the coulomb interaction for 
distances greater than half of the super-cell size. In the second one, ``WS cut'', the coulomb interaction is unchanged 
inside the Wigner-Seitz cell and periodically repeated outside. Both these two approximations converge fast and to the 
correct limit increasing the simulation box volume. In particular the Wigner-Seitz truncation is the most effective and 
already for a super-cell lattice size of $20$ bohr gives very well converged results for both RPA and RPAx correlation 
energies.

To be consistent, the same strategy has been applied to the calculation of EXX energy. 
The integrable divergence appearing in the reciprocal-space expression of the EXX energy in Eq.~\ref{eq1.2bis}, is a
well known problem and can be dealt with, for example, using the scheme proposed by Gygi
and Baldereschi~\cite{gygi_self-consistent_1986} (GB). 
Using this scheme one gets a convergence behavior (black circle in Fig.~\ref{fig:conv-Ex-alat}) that is proportional
to the inverse of the simulation box volume and eventually goes to the infinite volume limit represented by the dashed
black lines in the figure. An extrapolation scheme~\cite{nguyen_efficient_2009} can be used to mitigate this rather
slow convergence with respect to the simulation box volume (``GB+extrapolation'' in the figure). 
Also in this case, both modified coulomb interactions recover the correct limit
and in particular the Wigner-Seitz renormalization gives converged results within few tenths of meV already for a
super-cell lattice size of $a=25$ bohr.

\begin{figure}[t]
\begin{center}
\includegraphics[scale=0.3]{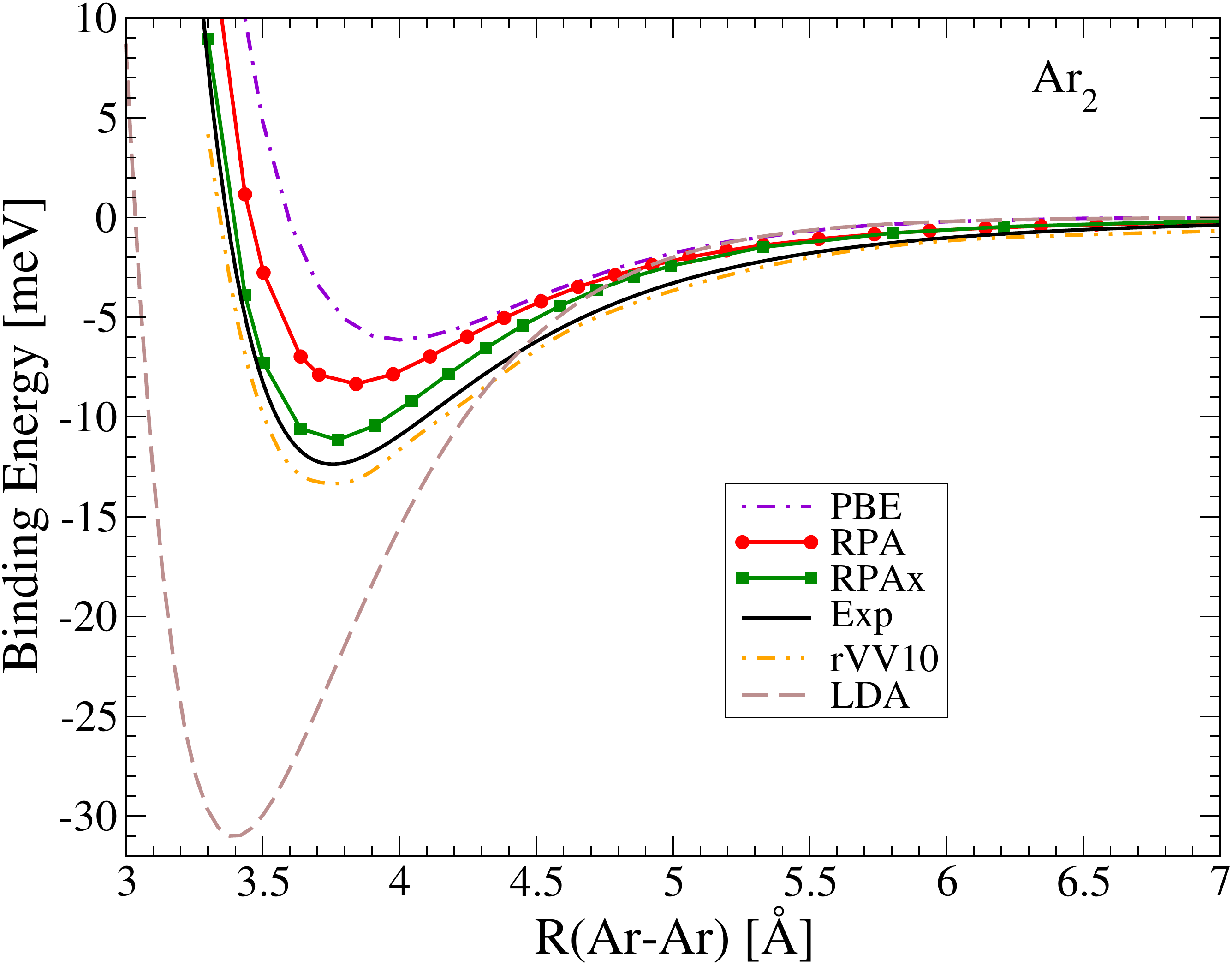}
\caption{Dissociation  curve of Ar$_2$ molecule. The plot compares results from PBE, RPA and
RPAx calculations.}
\label{fig:Ar2-diss}
\end{center}
\end{figure}

\section{Bond dissociation of dimers}\label{sect:res}

The correct description of chemical bonds of different nature is a fundamental prerequisite 
for any accurate electronic structure method. In the following we will consider diatomic 
molecules with different types of chemical bonds ranging from weak van der Waals bonds 
(Ar$_2$ and Kr$_2$) to mixed covalent-van der Waals bonds (Be$_2$ and Mg$_2$) to an 
ionic/covalent bond (LiH). We study the full potential energy curves which allows us to assess 
the general accuracy and applicability of the RPAx functional. 

\subsection{van der Waals dimers}

\begin{figure}[t]
\begin{center}
\includegraphics[scale=0.3]{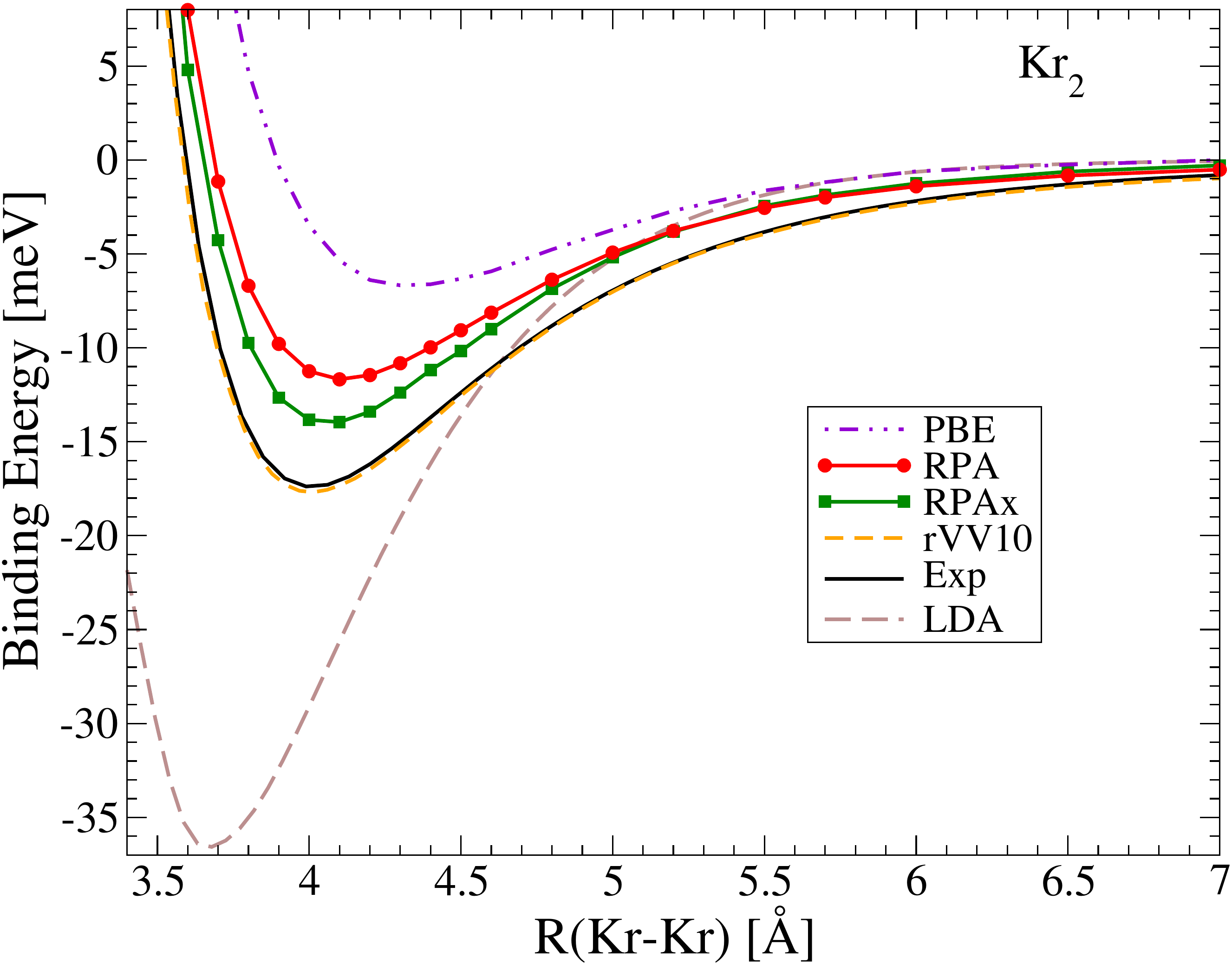}
\caption{Dissociation curve of Kr$_2$ molecule. The plot compares results from PBE, RPA and
RPAx calculations.}
\label{fig:Kr2-diss}
\end{center}
\end{figure}

Although one of the weakest, the dispersion (or van der Waals) interaction is of primary importance for 
an accurate understanding of, for instance, biological processes as well as adsorption on surfaces 
and chemical reactions. Dispersion forces originate from the response of an electron in one 
region to instantaneous charge-density fluctuations in another one. The leading contribution arises 
from the dipole--induced-dipole interaction and leads to an attractive energy with the well-known 
$-1/R^6$ decay with the inter-atomic separation $R$. Standard xc functionals give a contribution only 
when there is an overlap of density charges of the individual components of the system. Because 
the overlap decay exponentially with the inter-atomic separation so does the binding energy.

Correlation energy functionals derived within the ACFD formalism are fully non-local and therefore include automatically
and seamlessly dispersion interactions (see, for instance, Ref.~\onlinecite{dobson_quasi-local-density_2003} for a
detailed proof). They perfectly combine with the exact-exchange energy, thus removing the ambiguity, present in
almost any other approach to van der Waals systems, in the choice of how to combine exchange, short-range
correlation and non-local correlation contributions. 
\begin{table}[b]
\caption{\label{tab:struc-par-vdw} Equilibrium properties of argon and krypton
dimers computed within different functionals: LDA, PBE, rVV10, RPA and RPAx.  
Equilibrium bond length ($R_0$) in \AA\, binding energy ($E_b$) in meV ,
and vibrational frequency ($\omega_0$) in cm$^{-1}$. Experimental data from Ref.~\onlinecite{tang_van_2003} are also 
reported. }
\begin{ruledtabular}
\begin{tabular}{l l c c c c c c }
      & LDA & PBE & RPA & RPAx & rVV10 & Exp. \\
\hline
Ar$_2$ & & & & & \\ 
      $\quad R_0$(\AA) 		  & 3.39 & 3.99 & 3.84 & 3.75 & 3.75 & 3.76 \\
      $\quad E_b$(meV) 		  & 31.0  & 6.1  & 8.3  & 11.1  & 13.4  &  12.4  \\
      $\quad \omega_0$(cm$^{-1}$) & 58.3  & 23.4  & 26.8  & 30.8  & 32.3  &  31.2 \\
\hline
Kr$_2$ & & & & & \\
      $\quad R_0$(\AA) 		  & 3.67 & 4.33 & 4.11 & 4.06 & 4.01 & 4.01 \\
      $\quad E_b$(meV) 		  & 36.6  & 6.7  & 11.7  & 14.0 & 17.7  &  17.4  \\
      $\quad \omega_0$(cm$^{-1}$) & 39.8  & 15.4  & 20.1  & 22.8  & 23.8  &  23.6 \\
\end{tabular}
\end{ruledtabular}
\end {table}

Having binding energies (BEs) of the order of tens of meV, noble-gas dimers represent interesting test cases 
to investigate the accuracy of RPA and RPAx methods. Most of the RPA calculations for realistic systems in general 
have been performed in a non self-consistent-field (non-scf) fashion, namely,
exact-exchange and RPA correlation energies were computed using single particle orbitals obtained from a local or
semi-local self-consistent DFT calculation.
Only recently Nguyen \emph{et al}.~\cite{nguyen_ab_2014} have performed for the first time a fully scf calculation for
Ar$_2$ and Kr$_2$ at the RPA level revealing a close agreement between the scf-RPA and the RPA@PBE dissociation curves.
This indicates that the PBE density is rather close to the scf-RPA density and thus justifies the use of this density in
non-scf calculations instead of performing a full scf-RPA one. According to these finding and in absence of a scf-RPAx
method, we performed our RPAx correlation energy calculations starting from well converged PBE orbitals.

The dimers and the corresponding isolated atoms have been simulated using a simple-cubic super-cell with a linear size 
$a=25$ bohr. The electron ion interactions have been described by conventional norm-conserving 
pseudo-potentials~\cite{hamann_norm-conserving_1979}; a kinetic energy cut-off of $80$ Ry and $50$ Ry for Ar and Kr, 
respectively, has been used. Finally we used up to $400$ low-lying eigenvalues of the RPA/RPAx response function in
order to calculate the corresponding correlation energies. 
Extensive tests have been conducted to ensure that these parameters give well converged binding energy with errors
estimated to be less than 1meV.

In Fig.~\ref{fig:Ar2-diss} and Table~\ref{tab:struc-par-vdw} we report our results for the RPAx@PBE dissociation curve
of Ar dimer (green squares)
together with those of several DFA and compare them with an accurate model potential fitted on experimental
data~\cite{tang_van_2003} (black solid line). The RPAx and RPA binding energies have been calculated using well
converged PBE orbitals while all the other density functional calculations are fully self-consistent.
As expected LDA and PBE (and GGAs in general) give very poor results predicting either too large or too small binding 
energies and equilibrium distances.
Including the exact-exchange kernel leads to an overall improvement of the RPA performance. This results is 
consistent with the improved van der Waals coefficients obtained in Ref. \onlinecite{hellgren_correlation_2010}. 
Compared to the binding energy ($12.4$ meV), bond-length ($3.76$~\AA) and vibrational frequency ($31.2$ cm$^{-1}$) 
obtained from a model potential fitted to experimental data~\cite{tang_van_2003}, our RPAx results of $11.1$ meV, $3.75$~\AA\; and
$30.8$ cm$^{-1}$ show an impressive agreement. The RPAx dissociation curve turns out to be as good as the newly
developed vdW functional rVV10~\cite{sabatini_nonlocal_2013} which is a simple revision of the VV10 nonlocal density
functional by Vydrov and Van Voorhis~\cite{vydrov_nonlocal_2010}, specifically designed for van der Waals systems. 

Similar results are also observed for Krypton dimer and reported in Fig.~\ref{fig:Kr2-diss}.
Although comparison with dissociation energy curve obtained from an accurate model potential fitted to experimental
data~\cite{tang_van_2003} shows that RPAx scheme underestimate the binding energy by about $20$\%, the structural
properties at the RPAx level show an improvement if compared to the RPA. Our RPAx results for
the bond length and the vibrational frequency differ only by $1$\% and $-3$\% from the experimental values,
respectively, and compare better than RPA ($3$\%, $-15$\%) bond lengths and vibrational frequencies.
Only the rVV10 functional outshines the RPAx and gives result essentially identical to the experiments. 

\begin{figure}[t]
\begin{center}
\includegraphics[scale=0.3]{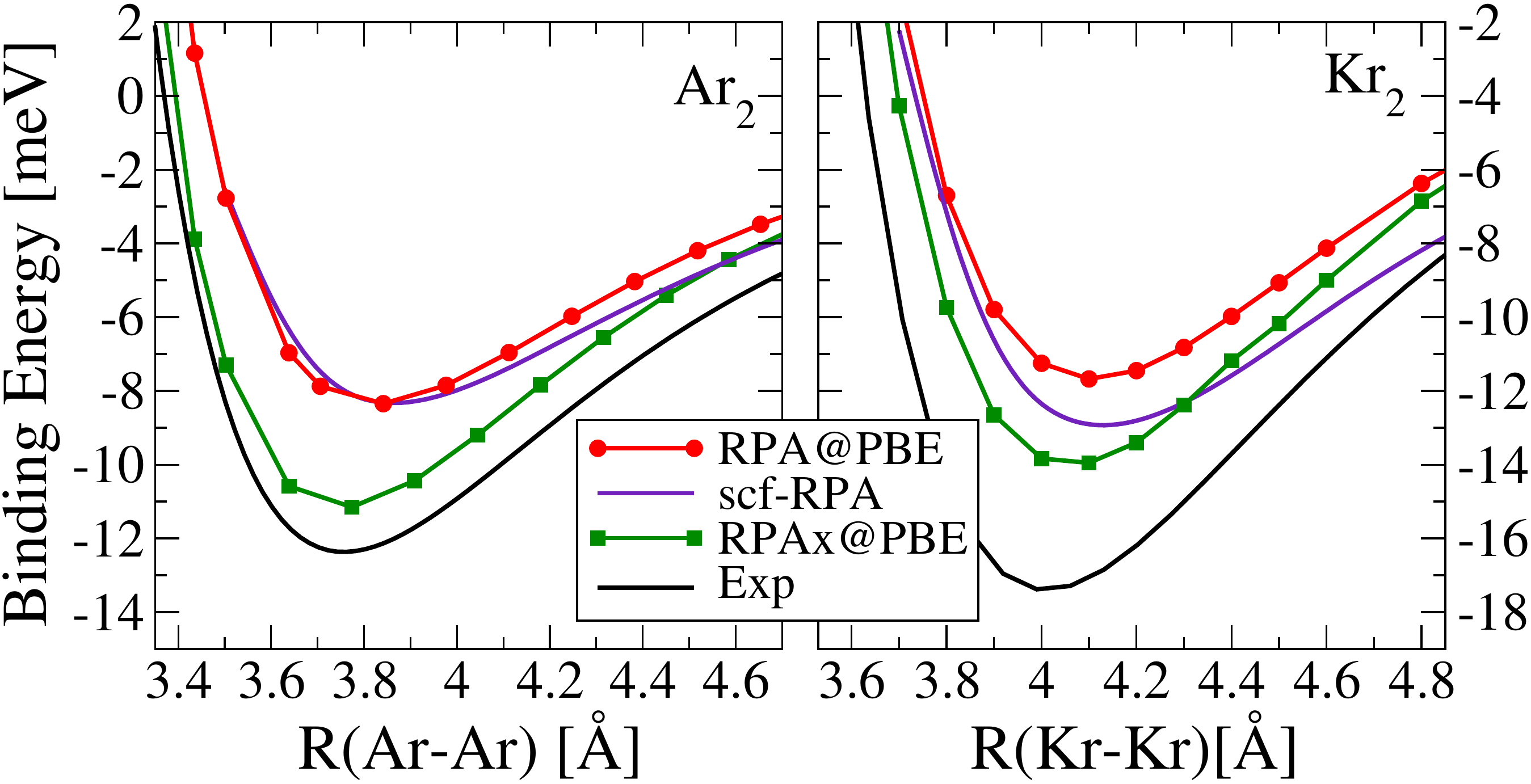}
\caption{Relative importance of the scf for Ar and Kr dimers.}
\label{fig:Ar2-Kr2-scfRPA}
\end{center}
\end{figure}

Despite the fact that Ar$_2$ and Kr$_2$ have very similar electronic structure, we notice a better
agreement with the experiment for Ar$_2$ than for Kr$_2$.
Fig.~\ref{fig:Ar2-Kr2-scfRPA} shows that the discrepancy can be
due in part to the different importance of the scf procedure in the two
systems. The left panel shows that the dissociation curves of the Ar$_2$
dimer computed at the RPA level using PBE or fully scf orbitals are very
close to each other, meaning that the PBE density is already a very good
approximation for the scf-RPA density.  On the other hand, passing from
the PBE density to the full scf-RPA one has a larger effect in the case
of Kr$_2$ dimer (right panel of Fig.~\ref{fig:Ar2-Kr2-scfRPA}) and a gain
of $\sim 1.3$ meV can be obtained in the binding energy. We therefore argue
a fully self-consistent RPAx calculation may lead to a similar gain in 
binding energy and ultimately to a very good agreement with the experimental 
results also for the Kr$_2$ case.

Similar accuracy for the Ar and Kr dimers has been found 
in Ref. \onlinecite{zhu_range-separated_2010} using a range separated hybrid 
RPAx approach (RSH-RPAx). The RPAx functional of that work is defined in terms of the 
Hartree-Fock kernel and evaluated with Hartree-Fock orbitals. In general, 
at full-range, this procedure worsens the results compared to RPAx defined 
within DFT as done in this work. However, by introducing the range separation 
results of similar quality can be obtained for van der Waals systems and even 
provide a better description of the Be dimer\cite{toulouse-09} which 
will be further discussed in the next section.

\subsection{Alkaline earth dimers}

\begin{figure}[t]
\begin{center}
\includegraphics[scale=0.3]{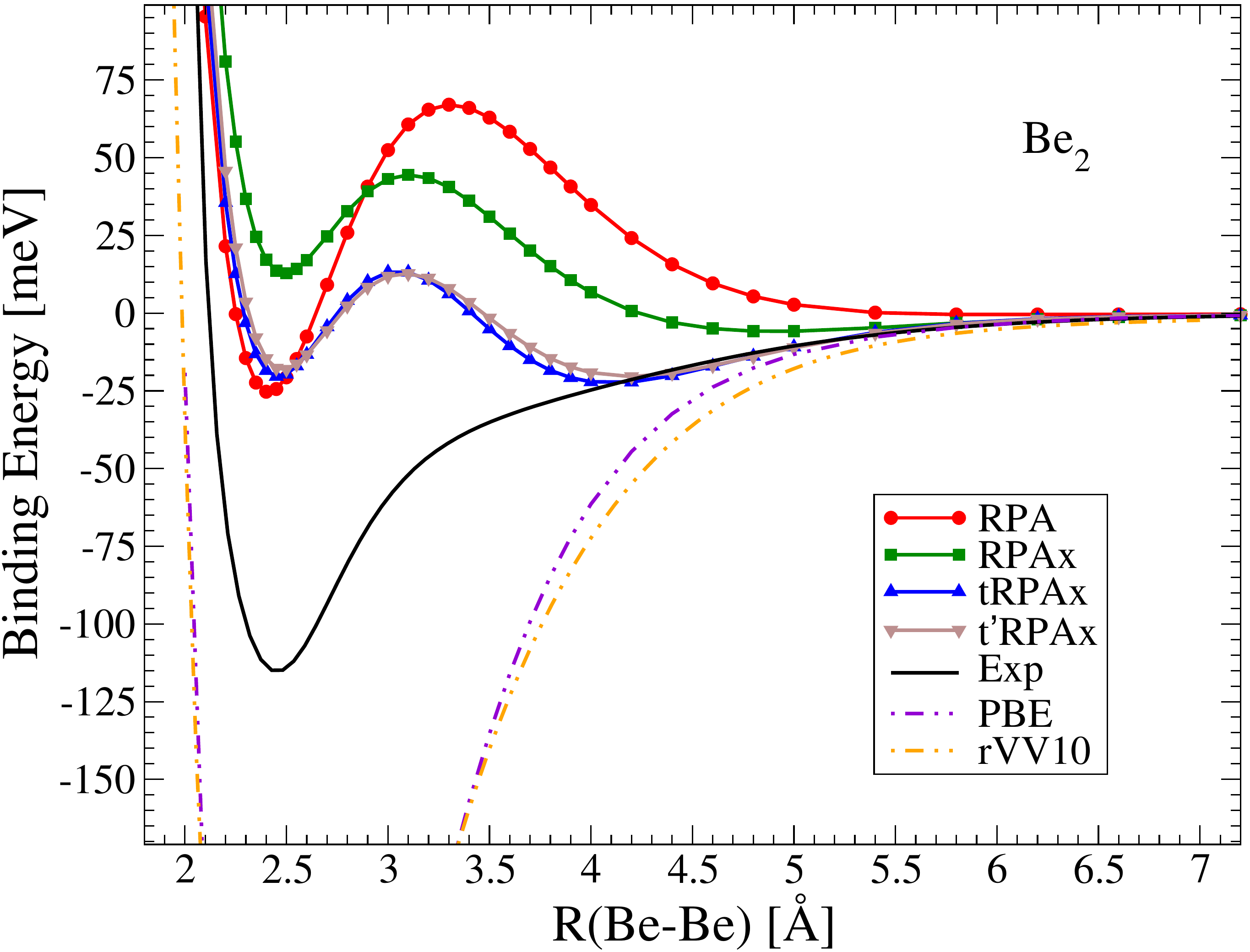}
\caption{Dissociation  curve of Be$_2$ molecule. The plot compares results from PBE, RPA and
RPAx calculations.}
\label{fig:Be2-BE}
\end{center}
\end{figure}

Alkaline earth dimers represent a much more complex situation than rare-gas dimers. Both long-range van der
Waals interaction and static correlation play an important role in these systems. They differ from the rare-gas 
atoms in having nearly degenerate $s-p$ sub-shells and hence multireference effects that are notoriously difficult to
capture by a single Slater determinant approach. 

Several theoretical investigations have been devoted to the simplest of these molecule, namely the beryllium dimer,
using different {\it ab initio} methods. Standard DFT calculations with local or semi-local functionals,
predict bond lengths rather close to the experimental value but binding energies about $500$\% too large (see, e.g.,
Ref.~\onlinecite{fuchs_accurate_2002}). Hartree-Fock or EXX does not even bind the molecule.\cite{nguyen_first-principles_2010,nguyen_ab_2014} 
Better results can be obtained resorting to Quantum Monte Carlo (QMC) techniques~\cite{marchi_resonating_2009, casula_diffusion_2005}
or high accuracy quantum chemistry methods such as the second-order M\o{}ller-Plesset perturbation
theory~\cite{lotrich_intermolecular_2005}, the couple-cluster approach~\cite{lotrich_intermolecular_2005} and 
the configuration interaction (CI) method~\cite{roeggen_interatomic_2005}. In addition, the study of 
dissociation energy curve and structure of Be dimer using RPA technique has also been carried 
out~\cite{fuchs_accurate_2002, toulouse-09,nguyen_first-principles_2010,nguyen_efficient_2009, nguyen_ab_2014}.
Although nscf-RPA can predict $R_0$ and $\omega_0$ in rather good agreement with experiment, the binding 
energy is severely underestimated. Moreover in Ref.~\onlinecite{nguyen_first-principles_2010} the presence of an unphysical
maximum in the nscf-RPA dissociation curve was found and the whole curve was shown to be very sensitive to the input 
orbitals used. The authors suggested a fully self-consistent treatment of the RPA density and potential as a possible
solution for the hump puzzle. Recently Nguyen \emph{et al.}~\cite{nguyen_ab_2014} have performed such a calculation for
this system and pointed out that the scf treatment is indeed important and significantly lower the total energy of the
system, yet it is not enough to fix the unphysical maximum problem and even leads to the metastability of the Be dimer.
These findings, as already pointed out in the original work, indicate a real limitation of the RPA that calls for the 
inclusion of correlation contributions beyond the simple Hartree kernel.

\begin{figure}[t]
\begin{center}
\includegraphics[scale=0.3]{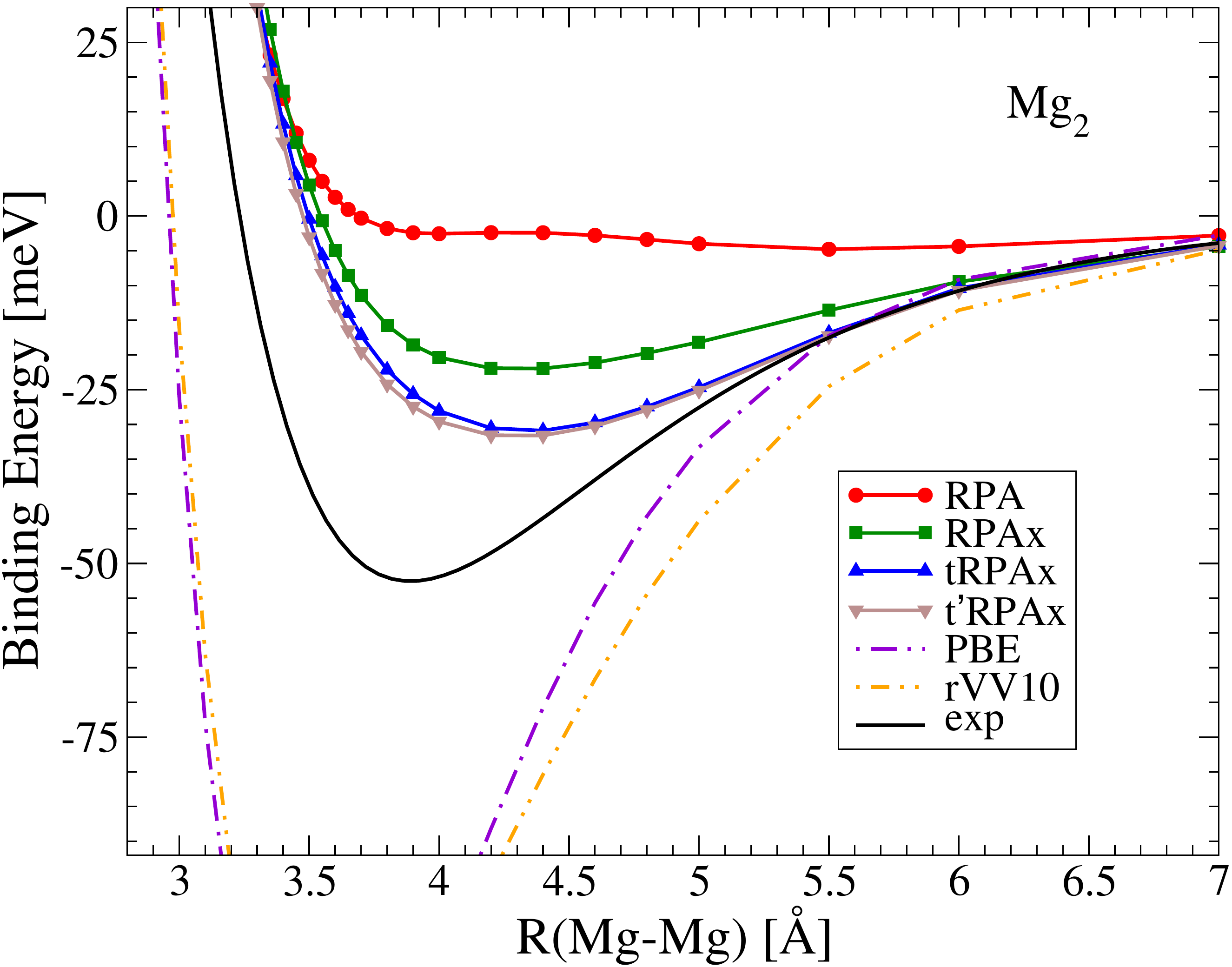}
\caption{Dissociation  curve of Mg$_2$ molecule. The plot compares results from PBE, RPA and
RPAx calculations.}
\label{fig:Mg2-BE}
\end{center}
\end{figure}
\begin{table}[b]
\caption{\label{tab:struc-par-alkaline} Equilibrium properties of beryllium and magnesium
dimers computed within different functionals: PBE, rVV10, RPA, RPAx and tRPAx.   
Equilibrium bond length ($R_0$) in \AA\, binding energy ($E_b$) in meV ,
and vibrational frequency ($\omega_0$) in cm$^{-1}$. Experimental data from Ref.~\onlinecite{merritt_beryllium_2009,balfour_absorption_1970} are also reported. }
\begin{ruledtabular}
\begin{tabular}{l l c c c c c c }
      & PBE & RPA & RPAx & tRPAx & rVV10 & Exp. \\
\hline
Be$_2$ & & & & & \\ 
      $\quad R_0$(\AA) 		  & 2.46  & 2.40  &  2.49  & 2.46 & 2.44 & 2.45 \\
      $\quad E_b$(meV) 		  & 399.4 & 25.1  & -13.1  & 20.0  & 387.7 & 115.3 \\
      $\quad \omega_0$(cm$^{-1}$) & 336   & 293   &  221   & 231  & 326 &  276 \\
\hline
Mg$_2$ & & & & & \\
      $\quad R_0$(\AA) 		  & 3.51  & -- & 4.31  & 4.30  &  3.56   &  3.89 \\
      $\quad E_b$(meV) 		  & 137.1 & -- & 22.3  & 30.9  &  131.1  &  52.57  \\
      $\quad \omega_0$(cm$^{-1}$) & 101.5 & -- & 27.0  & 31.2  &  90.2   &  51.1 \\
\end{tabular}
\end{ruledtabular}
\end {table}
In Fig.~\ref{fig:Be2-BE} and Table~\ref{tab:struc-par-alkaline} we report our results for the binding energy curve of Be$_2$ 
from standard PBE, rVV10, RPA and RPAx calculations and compare them 
with an accurate model potential-energy-surface fitted on experimental data~\cite{merritt_beryllium_2009}.
The results from the simple modifications (tRPAx and t$^{\rm \prime}$RPAx) of the original RPAx
proposed to fix its pathological behavior in the limit of small gaps~\cite{colonna_correlation_2014}, are also
shown. 
The dimer and the corresponding isolated atom have been simulated 
using a simple-cubic super cell with a linear size $a=25$  
bohr and a kinetic energy cut-off of $40$ Ry. Up to 200 lowest-lying eigenpairs of the 
generalized-eigenvalue problem in Eq.~(\ref{eq:RPAx-eig-PW}) have been used
to compute the RPA and RPAx correlation energies. All the RPA and RPAx
calculations have been performed starting from well converged PBE orbitals.

The PBE calculation predicts a bond length rather close to the experimental value but 
too large a binding energy ($400\%$ overestimate). The rVV10 functional that 
gave a very good description for purely van-der-Waals compound performs as poorly as
the PBE functional in this case. At the RPA level the dissociation 
curve exhibits an unphysical hump for intermediate values of the bond length very 
similar to the one observed for covalent bonded systems.~\cite{furche_molecular_2001, colonna_correlation_2014,
caruso_bond_2013} 
The RPA severely underestimates the binding energy but still gives equilibrium bond length and 
vibrational frequency in good agreement with experimental data. Passing from RPA to RPAx leads 
to a worse description of the dissociation curve near
the equilibrium position and even to the metastability of the Be dimer. However we 
notice that in the dissociation region the RPAx curve approaches the experimental 
potential energy surface faster than the RPA one in agreement with the largely improved 
description of the van der Waals forces within RPAx.\cite{hellgren_correlation_2010} 
The results obtained from the alternative re-summations tRPAx and t$^{\rm \prime}$RPAx 
defined in Ref. \onlinecite{colonna_correlation_2014}, recover the RPA performance 
near the minimum, reduce the height of the hump for intermediate values of the bond length 
and approach the correct asymptotic behavior of the Be-Be interaction potential much faster than 
all the other approximations.

The hump problem has been previously analysed in the literature, but is not yet fully understood. 
A spurious maximum in the exchange energy as a function of atomic separation appears when 
the EXX energy (Eq.~(\ref{eq1.2bis})) is evaluated with orbitals coming from a local potential,\cite{nguyen_first-principles_2010}
which partly explains the hump in the total energy. The use of self-consistent EXX orbitals reduces the total 
energy hump but does not completely remove it,\cite{nguyen_ab_2014} while the use of HF or hybrid orbitals 
generated by a non-local potential is sufficient to remove the hump for both RPA and RPAx dissociation curves 
\cite{nguyen_first-principles_2010,toulouse-09} but produces a very shallow minimum at large interatomic distance.
No energy hump and an underestimated binding energy is obtained for Be$_2$ in the recently developed r2PT 
approach~\cite{ren_renormalized_2013} that otherwise poorly describes the dissociation curve of H$_2$, that is 
instead satisfactorily described by RPAx.~\cite{colonna_correlation_2014}

The same basic perturbative ingredients, consistently treated in RPAx via G\"orling-Levy perturbation theory, are 
treated in r2PT starting from a Rayleigh-Schr\"odinger perturbation theory perspective, leading to the need to add 
single-excitation correction terms~\cite{rSE2011} to account for the inconsistency between the DFT reference and the 
HF-like perturbative 
approach used.~\cite{SE-RPAx} 
Higher order contributions from selected classes of energy diagrams amenable to resummation are also included.
RPAx and r2PT are therefore expected to contain very similar physical effects and the difference in performance of the two methods 
springs from the different approximate treatment of the high order contributions to the correlation energy.

As a second example of alkaline-earth dimer we studied the Mg$_2$ molecule. 
The dimer and the corresponding isolated atom have been simulated 
using a simple-cubic super cell with a linear size $a=25$  
bohr and a kinetic energy cut-off of $35$ Ry. Up to 180 lowest-lying eigenpairs of the 
generalized-eigenvalue problem in Eq.~(\ref{eq:RPAx-eig-PW}) have been used
to compute the RPA and RPAx correlation energies. All the RPA and RPAx
calculations have been performed starting from well converged PBE orbitals.

Our results for the PBE, rVV10, RPA and RPAx dissociation curves are shown in Fig.~\ref{fig:Mg2-BE} 
and Table~\ref{tab:struc-par-alkaline}, together with a model potential fitted on spectroscopic data~\cite{balfour_absorption_1970}. 
As already seen in the Be$_2$ molecule, also in this case the PBE functional greatly overbinds
the system still giving a reasonable value for the equilibrium distance. The rVV10 functional 
slightly modifies the PBE curve for large inter-atomic distances with no significant 
improvement in the equilibrium region. The RPA curve still exhibits a hump at intermediate distances, 
although much less pronounced than in the case of Be$_2$. As a result an almost flat curve is obtained 
for intermediate and large inter-atomic separations with a very shallow global minimum appearing at 
distances much greater than the experimental equilibrium one. 
Including the exchange contribution to the kernel improves the van der Waals forces which leads to a 
qualitative improvement in the description of the bond. The bump is no longer visible and a proper 
behavior for the dissociation curve is obtained with one minimum located in between the PBE and the RPA
one. The larger improvement within RPAx for the Mg$_2$ bond compared to the Be$_2$ bond suggest that 
this bond has a larger van der Waals character. Similarly to the Be molecule the alternative 
re-summations produce a better agreement with experiment.  
\begin{figure}[b]
\begin{center}
\includegraphics[width=8.6cm]{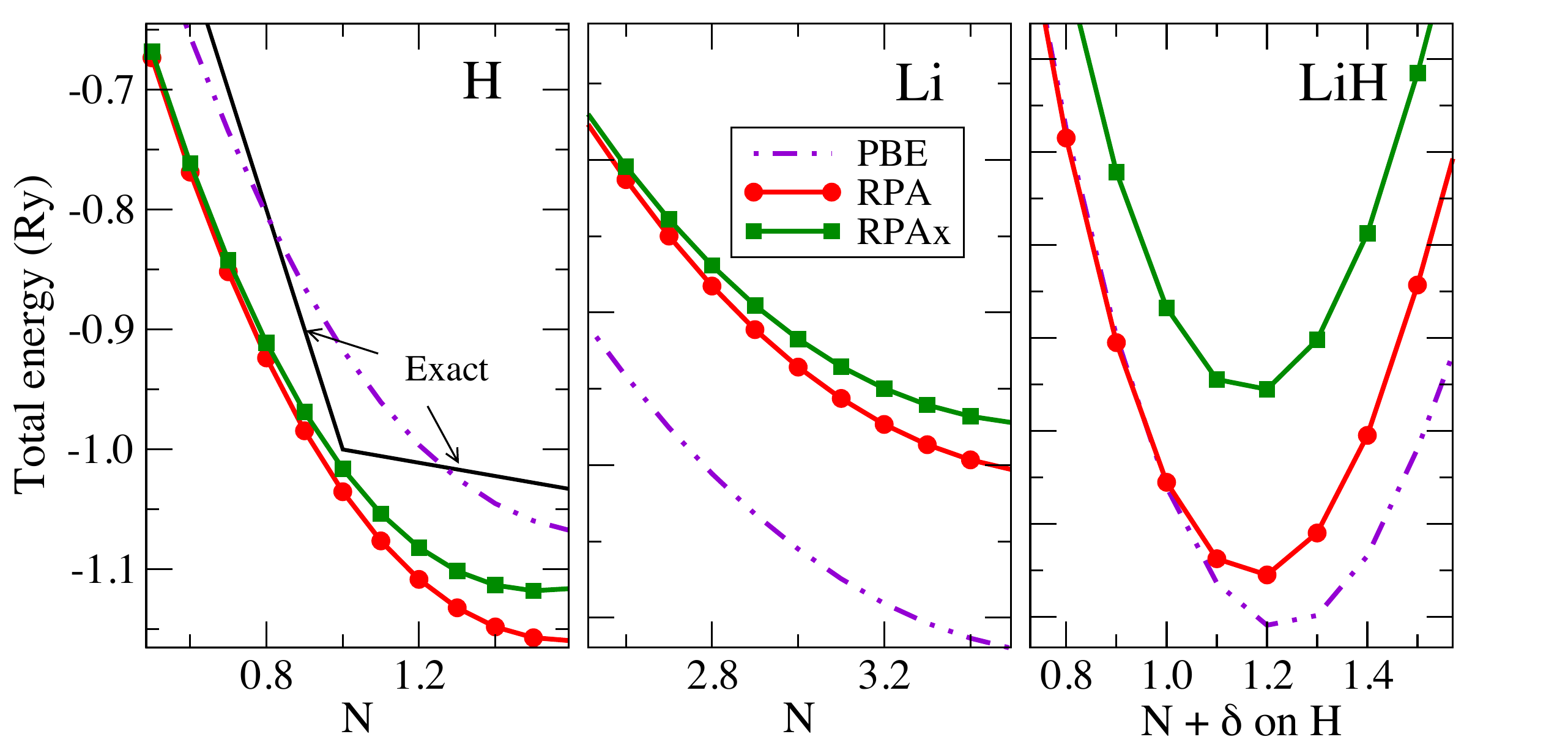}
\caption{Left and middle panels: Fractional charge analysis of the H and Li atoms. Right panel: 
Total energy of the LiH molecule as a function of fractional charge on the H atom.}
\label{fig:LiH-frac}
\end{center}
\end{figure}

\subsection{The LiH molecule}
The stretched H$_2$ molecule with its single covalent bond represents the simplest example 
of a strongly correlated system. Standard DFAs such as the LDA or GGAs fail but also more sophisticated 
approaches such as the MP2\cite{dahlen_variational_2006} and the GW approximation.\cite{hellgren_static_2015} 
A good total energy description is usually recovered resorting to an unrestricted calculation,
however at the price of having the wrong spin symmetry.
On the contrary, both RPA and RPAx are able to capture almost the full dissociation curve, as 
first shown in Ref.~\onlinecite{fuchs_describing_2005}. More recently, it was shown that also 
the strong triple bond of the N$_2$ dimer can be accurately captured by the tRPAx,\cite{colonna_correlation_2014} 
for which only CCSD(T) has been applied successfully so far.\cite{kowalski_N2} 
The common difficulty of H$_2$ and N$_2$ is the fact that they are composed of atoms with open shells, 
which results in a nearly degenerate situation as the bond is stretched. A similar situation occurs 
in the case of the LiH molecule which is described by a covalent/ionic bond. 
In addition, LiH is hetero-nuclear and as such the problem of charge delocalisation, as predicted by 
most functionals, is largely enhanced. As the LiH bond breaks the electrons should, in an 
exact treatment, localize and form two neutral atoms. However, standard approximations allow the 
electrons to delocalise and form fractionally charged dissociation fragments. This failure of 
common DFAs was first discussed in Ref.~\onlinecite{perdew_density-functional_1982} where it was linked to a missing 
derivative discontinuity of the xc functional and an associated step in the corresponding xc potential. 

In order to estimate this type of delocalisation error one can determine
the amount of fractional charge a given functional predicts by means of a fractional 
charge analysis.\cite{cohen_insights_2008} The functional is extended to allow for densities that 
integrate to a non-integer particle number and the energy of the independent 
atoms are calculated as a function of particle number. The total energy of the combined
system (i.e. the molecule at infinite separation) can then be determined as a function
of the amount of charge on e.g. the H atom and the minimum can be located. 
\begin{figure}[t]
\begin{center}
\includegraphics[scale=0.3]{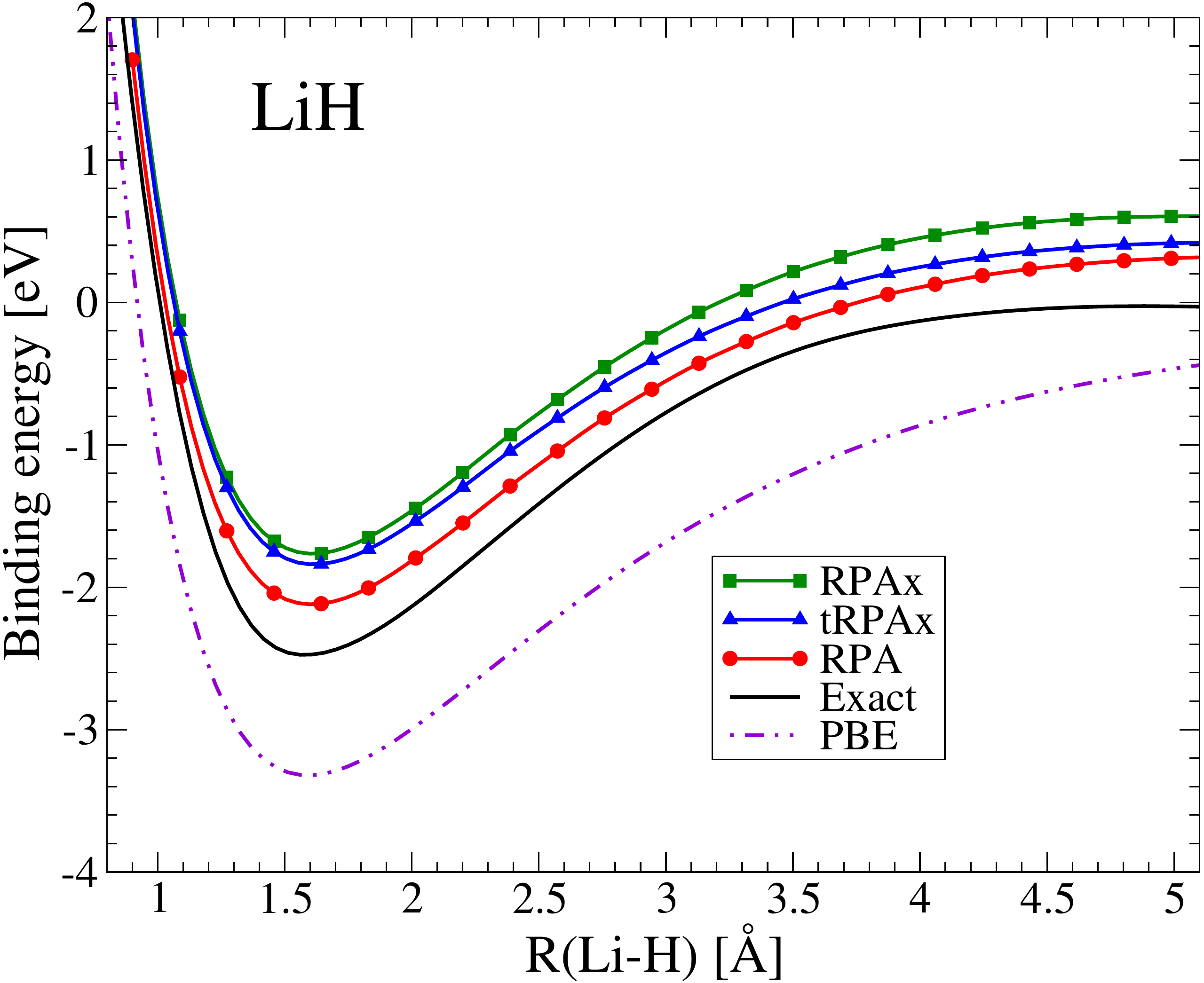}
\caption{Binding energy curve of the LiH molecule. The plot compares results from PBE, RPA,
RPAx and tRPAx calculations.}
\label{fig:LiH-BE}
\end{center}
\end{figure}
\begin{figure*}[t]
\begin{center}
\includegraphics[scale=0.3]{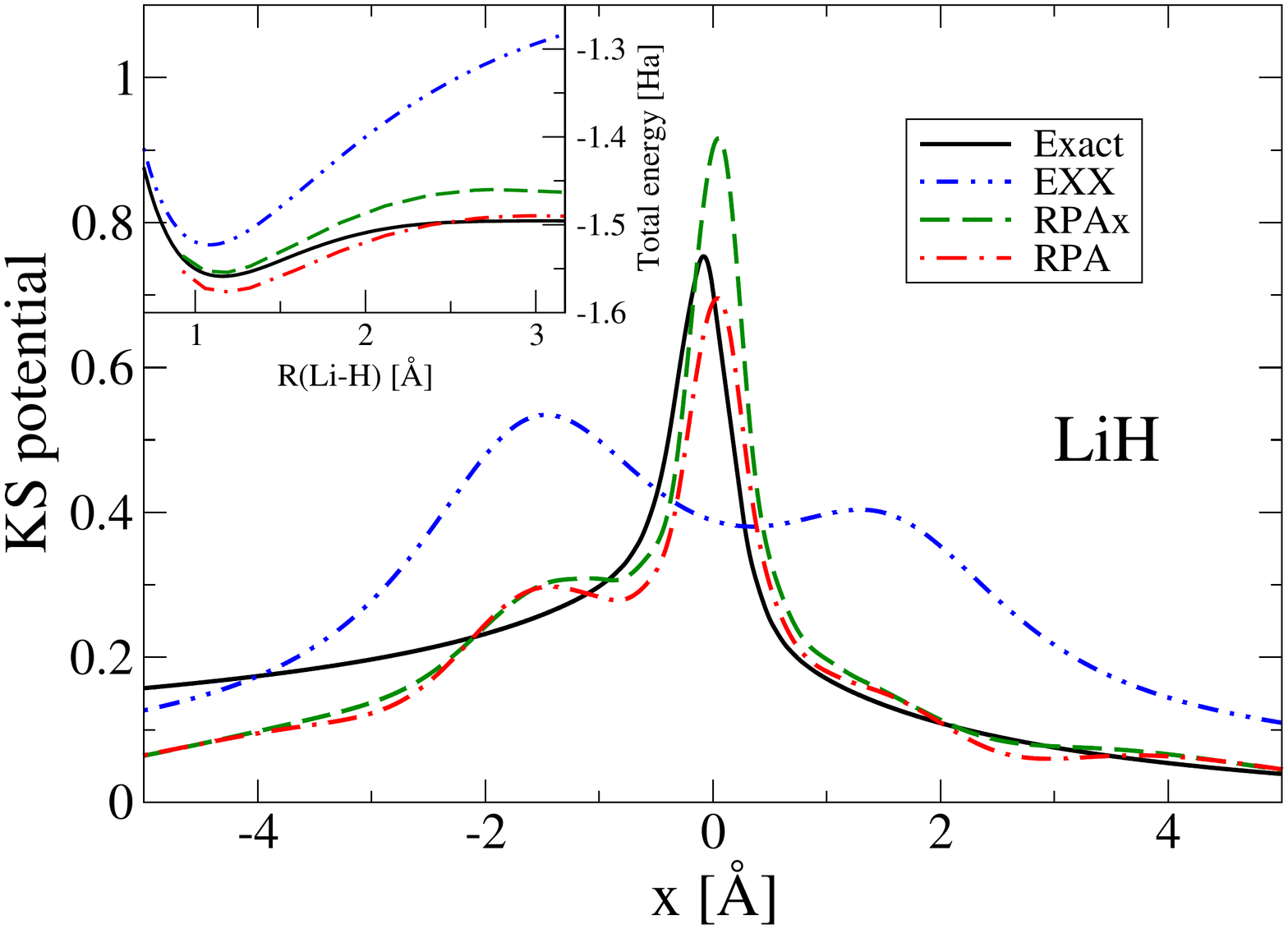}
\includegraphics[scale=0.3]{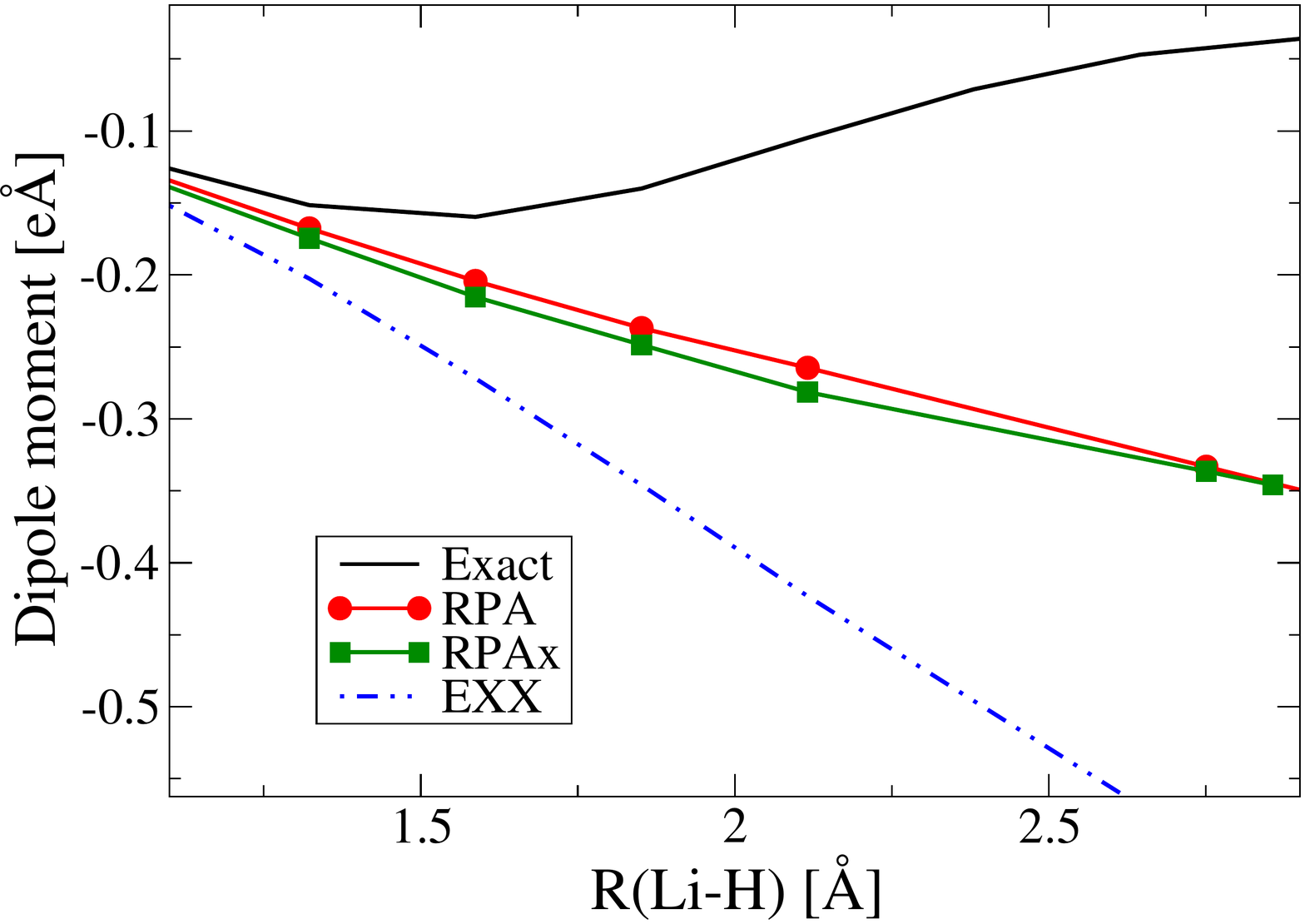}
\caption{(Left) The KS potentials of a model LiH at $R=3.5$\AA. $x=0$ corresponds to the bond midpoint. Inset 
shows the corresponding potential energy curves. (Right) The dipole moment as a function of nuclear separation of the 
exact density and the self-consistent densities within EXX, RPA and RPAx.}
\label{fig:LiH-pot}
\end{center}
\end{figure*}
We carried out such a fractional charge analysis for LiH using the PBE, RPA and RPAx functionals. 
The results can be found in Fig.~\ref{fig:LiH-frac}. Notice that the pseudo potential 
approximation has been applied to the Li-atom (middle panel of Fig.~\ref{fig:LiH-frac}) 
so we cannot compare the total energy to the exact result. This can, however, be done 
in the case of the H-atom, see left panel of Fig.~\ref{fig:LiH-frac}. 
As expected, at integer particle number RPA and RPAx outperform the PBE. However, we also see 
that as soon as we deviate from an integer both RPA and RPAx are smooth with a rather 
large curvature. As already found within the RPA~\cite{hellgren_correlation_2012}
also within the RPAx the derivative discontinuity is missing meaning that the inclusion 
of the exchange kernel does not improve the qualitative behavior of the RPA for fractional 
particle numbers. In the third panel we see that the RPAx minimizes at 1.2 electrons on the 
H-atom which is a result comparable to those of the RPA and PBE. This result is perhaps not 
surprising since for two electrons the EXX kernel is Hartree-like and thus does not 
introduce any new correlation effects. If the RPAx calculation for 
the LiH molecule was performed self-consistently we would obtain 1.2 electrons on the 
H-atom in the dissociation limit. To be consistent when 
calculating the binding energy we should therefore subtract the value of the energy at this 
minimum and not the energy of the two atoms with integer particle numbers. In Fig.~\ref{fig:LiH-BE} 
we have plotted the binding energy curve of LiH with the energy in RPA and RPAx evaluated 
using PBE orbitals. Since the PBE functional minimizes at a value close to RPA(RPAx) we expect the
result to be close to the self-consistent result. We see that, although RPAx gives a better total
energy (not shown), the binding energy is slightly worse than RPA, still gives an improvement 
compared to PBE. The use of tRPAx does not improve the results much. The fact that RPA and RPAx
perform worse for LiH as compared to the H$_2$ is a result of the additional delocalisation 
error. We also notice a very slow convergence to the dissociation limit. The maximum of the hump 
has not been reached at 5 \AA. This slow convergence can also be attributed to the delocalisation 
error as we will now further discuss.

Recently, fully self-consistent RPA was implemented within {\sc Quantum ESPRESSO}.\cite{nguyen_ab_2014} 
However, in order to compare with exact results we decided to simplify the problem and model the 
LiH molecule as a one-dimensional system with a softened Coulomb interaction. For this problem we 
could use an implementation based on cubic splines as basis functions which is known to solve 
the optimized effective potential equation with high precision.\cite{hellgren_correlation_2007,hellgren_correlation_2012} 
To obtain the exact potential we used {\sc OCTOPUS}.\cite{octopus} 
In Fig.~\ref{fig:LiH-pot} (left) we plot the full KS potential at a bond distance of $R=3.5$ \AA~
in RPA, RPAx and EXX and compare the results to the exact potential. The bond midpoint is located at $x=0$ and 
the corresponding potential energy curves are shown in the inset. The RPA is in rather good agreement
with the exact potential. It exhibits the correlation peak at the bond midpoint but it misses the step
feature. The RPAx is similar to RPA but with an overestimated bond midpoint peak. Again the step 
feature is missing which is consistent with the missing derivative discontinuity in the xc energy. 
The large humps of the EXX potential is a result of the large delocalisation error in EXX which has 
been estimated to 1.4 electrons on the H-atom in previous works.\cite{perdew_density-functional_1982,hellgren_static_2015} 

In order to understand the huge impact of the missing step or derivative discontinuity on the density
we have also calculated the dipole moment of the LiH molecule, see Fig. ~\ref{fig:LiH-pot} (right).
In the exact treatment the dipole moment should tend to zero in the dissociation limit since the molecule
dissociates into neutral atoms. Due to the fact that approximate functionals produce fractional 
charges on the dissociation fragments the dipole moment will instead tend to infinity.\cite{hellgren_static_2015} 
Indeed, this is exactly what we see for the approximations studied and the bond breaking transition 
is missing in the self-consistent density of the EXX, RPA and RPAx.

\section{Conclusions}\label {sect:summ}

We presented an assessment of RPA and RPAx functionals on a set of diatomic molecules
having chemical bonds of different nature. For weakly interacting systems we found that the inclusion
of the exact-exchange kernel leads to a quantitative improvement with respect to the RPA results.
The binding energy and the structural properties of Ar$_2$ and Kr$_2$ molecules show a significant 
improvement when moving from the RPA to the RPAx approximation due to an improved description of the van
der Waals forces.
For the binding energy curve of Be$_2$ even if a better description of the dissociation region and a 
reduction of the hump at intermediate inter-atomic distances can be achieved within the tRPAx approximation,
the binding energy is still far from the experimental results and a potential energy surface with 
two minima is obtained. On the other hand, the dissociation curve of the Mg$_2$ molecule shows a 
much larger improvement passing from the RPA to the RPAx approximation due to the increased van 
der Waals character of the bond. The double minimum structure found at the RPA level 
is correctly removed by including the exact-exchange contribution to the kernel and a well behaved potential
energy surface is recovered. A comparison between results for purely van der Waals and mixed covalent-van der Waals 
molecules reveals the difficulty of RPA and RPAx functionals in describing situations where the energetic and structural
properties of the system result from a delicate balance between bonding forces of different nature. Nevertheless
also in these difficult cases the long range part of the dissociation curves display a sensible improvement passing
from the RPA to the RPAx. Finally, we addressed the problem of strong correlation by analysing the 
dissociation of the LiH molecule. The RPA gives a qualitative improvement of the total energy as 
compared to standard density functionals. The inclusion of the exchange kernel within RPAx further corrects the 
correlation energy and gives results in good agreement with experiment. The hump at intermediate bond 
distances remains, however, and is shown to be a symptom of a large error in the corresponding RPA/RPAx densities. 
This error originates from a poor description of fractional charges and a missing derivative discontinuity 
within RPA and RPAx.

Although in the present work we focused on molecular systems, for which exact reference data are 
available for comparison, the current approach can easily be extended to solids and surfaces.
The additional computational workload required by the Brillouin zone
integration (see eq.~9) should be largely compensated by the reduced 
number of plane-waves associated to a much smaller unit cell. We 
therefore believe the application of the method to extended systems
to be as computational demanding as the calculations presented here.  
Work toward this goal is underway.

\begin{acknowledgments}
NC thanks Dr. Ngoc Linh Nguyen for useful discussions.
We thank CINECA (Bologna, Italy) and SISSA (Trieste, Italy) for the availability 
of high performance computing resources and support. 
Work supported by the Italian MIUR through the PRIN 2010 initiative (PRIN 20105ZZTSE).

\end{acknowledgments}

\bibliographystyle{aip}

\end{document}